\documentclass[12pt]{article}

\usepackage[dvips]{epsfig}
\usepackage[dvips]{graphics}
\input{psfig.sty}

\def\be{\begin{equation}}
\def\ee{\end{equation}}
\def\bea{\begin{eqnarray}}
\def\eea{\end{eqnarray}}

\newtheorem{theorem}{Theorem}[section]

\newcommand{\firstG}{\mathcal{G}_1}
\newcommand{\secondG}{\mathcal{G}_2}
\newcommand{\thirdG}{\mathcal{G}_3}
\newcommand{\fourthG}{\mathcal{G}_4}
\newcommand{\fifthG}{\mathcal{G}_5}
\newcommand{\sixthG}{\mathcal{G}_6}
\newcommand{\firstB}{\mathcal{B}_1}
\newcommand{\secondB}{\mathcal{B}_2}
\newcommand{\thirdB}{\mathcal{B}_3}
\newcommand{\fourthB}{\mathcal{B}_4}

\title{\bf Detectability of cosmic topology in f\/lat 
                 universes} 

\author{
G.I. Gomero\thanks{german@cbpf.br} 
\ and \  
M.J. Rebou\c{c}as\thanks{reboucas@cbpf.br}, \\ 
\\
Centro Brasileiro de Pesquisas F\'\i sicas, \\ 
Rua Dr. Xavier Sigaud 150 \\
22290-180, Rio de Janeiro -- RJ, Brazil
}

\begin{document}

\date{\today}

\maketitle

\begin{abstract} \noindent
Recent observations seem to indicate that we live in a 
universe whose spatial sections are nearly or exactly f\/lat.
Motivated by this we study the problem of observational 
detection of the topology of  universes with f\/lat spatial 
sections. We f\/irst give  a complete description of the 
dif\/feomorphic classif\/ication of compact f\/lat $3$-manifolds, 
and derive the expressions for the injectivity radii, 
and for the volume of each class of Euclidean $3$-manifolds. 
There emerges from our calculations the undetectability 
conditions for each (topological) class of f\/lat universes. 
To illustrate the detectability of f\/lat topologies we 
construct toy models by using an assumption by 
Bernshte\v{\i}n and Shvartsman which permits to establish a 
relation between topological typical lengths to the dynamics 
of f\/lat models. 
\end{abstract}
\newpage

\section{Introduction}

A great deal of work has recently gone into studying the 
possibility that the universe may possess compact spatial 
sections with a nontrivial topology, including the 
construction of dif\/ferent topological indicators 
(see, for example, refs.~\cite{CosmicTop}~--~\cite{ZelNov83}).
A fair number of these studies have concentrated on cases 
where the densities corresponding to matter and vacuum energy 
are substantially smaller than the critical density. This 
was motivated by the fact that until very recently observations 
favoured a low density universe. However, recent measurements 
of the position of the f\/irst acoustic peak in the angular 
power spectrum of cosmic microwave background radiation (CMBR) 
anisotropies, by BOOMERANG-98, MAXIMA-I, and WMAP experiments, 
seem to provide strong evidence that the corresponding ratio for 
the total density to the critical density, $\Omega_0$, is 
close to one~\cite{BooMax}~--~\cite{WMAP}.

In addition an important aim of most of previous works in Cosmic 
Topology has often been to produce examples where the topology 
of the universe has strong observational signals, and can 
therefore be detected and even determined. Until very recently 
it was never considered in detail the possibility that the 
topology of the universe may not be detectable from the current 
astro-cosmological observations due to its almost 
\emph{flatness}. 

In two previous articles~\cite{GRT1,GRT2} we have studied the 
question of detectability of a possible nontrivial compact 
topology in locally homogeneous and isotropic universes with 
total density parameter close to (but dif\/ferent from) one, 
i.e. the so-called \emph{nearly flat hyperbolic or spherical} 
universes. We have employed an indicator $T_{inj}$ which is 
def\/ined by the ratio of the \emph{injectivity radius}, 
$r_{inj}\,$, to the depth of a given catalogue, $d_{obs}$. 
In recent articles Gausmann \emph{et al.\/}~\cite{EvLeLuUzWe}
and Weeks~\cite{WeLeUz} \emph{et al.\/} have also studied which spherical 
topologies are likely to be detectable by using crystallographic 
and circles-in-the sky methods. 
More recently Gomero \emph{et al.\/}~\cite{GRT3} and Weeks~\cite{Weeks2002}
have discussed the detectability of cosmic topology of nearly f\/lat
hyperbolic universes. An important 
outcome from these studies~\cite{GRT1}~--~\cite{Weeks2002} 
is that by using any method of detection of topology which 
rely on observations of repeated patterns the topology of an 
increasing number of nearly f\/lat (hyperbolic and spherical) 
becomes undetectable as $\Omega_0 \to 1$. Thus, it would appear 
at f\/irst sight that in the limiting case $\Omega_0 = 1$ the 
topology of FLRW universes would def\/initely be undetectable. 
However, it is well known that when $\Omega_0$ is exactly one 
the topological possibilities for the universe are completely 
dif\/ferent, and the detectability of cosmic topology may again 
become possible. In this article we complete our previous 
works~\cite{GRT1,GRT2} by extending the analysis of detectability 
of cosmic topology to the f\/lat cases, by considering multiple
images of clusters of galaxies or pattern repetitions in CMBR
maps. 

The underlying cosmological setting and new results of this 
paper are stated and structured as follows. In 
Section~\ref{CosSet} we give an account of the cosmological 
framework employed throughout this work. Section~\ref{Flat3Man} 
gives a complete description of the dif\/feomorphic 
classif\/ication of compact f\/lat $3$-manifolds. In this 
section we also derive the expressions for the injectivity 
radii $\,r_{inj}\,$ and present the formulae (derived in 
Appendix~B ) for the volume for each class of Euclidean compact 
$3$-manifolds. In Section~\ref{DetTop} we present a brief 
discussion of the question of detectability of topology, 
recasting (and ref\/ining upon) some of the detectability 
aspects discussed in our previous articles~\cite{GRT1,GRT2}. 
In Section~\ref{Toys} we use the results of Sections~\ref{Flat3Man} 
and~\ref{DetTop} in connection with an assumption f\/irst 
suggested by Bernshte\v{\i}n and Shvartsman~\cite{BerShvar} to 
illustrate the problem of the detectability of f\/lat topologies. 
We remark from the outset that our use of an \emph{ad hoc} hypothesis, 
such as that of Bernshte\v{\i}n and Shvartsman's, is just for 
illustrative purposes, and we are not suggesting that this particular 
hypothesis is a realistic assumption for establishing a relation 
between topological typical lengths and the dynamics of f\/lat 
models. As a particular result we show that all models of two 
specif\/ic classes of f\/lat universes which satisfy 
Bernshte\v{\i}n-Shvartsman hypothesis have a \emph{detectable} 
topology (at least in principle for some observers) if the 
existing catalogues of clusters of galaxies are used. However, 
we also show that an alternative version of 
Bernshte\v{\i}n-Shvartsman assumption leads to f\/lat 
universes with \emph{undetectable} topologies, even if CMBR is 
used.  Section~\ref{Concl} contains a summary of our main results 
and further remarks. The details of the isometric classif\/ication 
of compact f\/lat $3$-dimensional manifolds is treated in 
Appendix~A. Finally in Appendix~B we  
present the relevant piece of calculations which lead to the 
expressions of the volume of all compact Euclidean $3$-manifolds.

\section{Cosmological setting} 
\label{CosSet}

We shall assume in this work that the universe is modelled by 
a $4$-manifold $\mathcal{M}$ which allows a $(1+3)$ splitting, 
$\mathcal{M} = R \times M$, endowed with Robertson-Walker metric 
with locally f\/lat spatial sections
\be
\label{FLRW1}
ds^2 = -c^2dt^2 + R^2 (t) \left [ d \chi^2 + \chi^2 
(d\theta^2 + \sin^2 \! \theta \,  d\phi^2) \right ]\; ,
\ee
where $t$ is a cosmic time, $c$ is the speed of light, and 
$R(t)$ is the scale factor. Furthermore, we shall also assume 
that the $3$-space $M$ is a compact f\/lat manifold, i.e. 
$M = E^3/\Gamma$, where $\Gamma$ is a discrete group of 
Euclidean isometries acting freely on $E^3$. 
The group $\Gamma$ is the so-called the covering group of 
$M$, and is isomorphic to the fundamental group $\pi_1(M)$.

For non-f\/lat models ($k \neq 0$), the scale factor $R(t)$ is 
identif\/ied with the curvature radius of the spatial section 
of the universe at time $t$, and thus $\chi$ can be interpreted 
as the distance of any point with coordinates $(\chi, \theta, 
\phi)$ to the origin of coordinates (in the covering space) in 
units of curvature radius, which is a natural unit of length. For 
f\/lat models, $\chi$ can still be interpreted as the distance of 
any point with coordinates $(\chi, \theta, \phi)$ to the origin of 
coordinates in units of $R(t)$, but since the curvature radius of 
Euclidean $3$-space is inf\/inite, one cannot identify $R(t)$ with 
the curvature radius, and so in this case there is no natural unit 
of length. Thus in this work, dif\/ferently from refs.~\cite{GRT1,GRT2}, 
we will use megaparsecs ($Mpc$)as unit of length.

In the light of current observations, we assume the current 
matter content of the universe to be well approximated by 
pressureless dust plus a cosmological constant~\cite{Lambda}. 
The redshift-distance relation in FLRW models with Euclidean 
spatial sections can be written in the form 
\begin{equation}
\label{redshift-dist}
d(z) = \frac{c}{H_0} \int_1^{1+z} \!\!\! 
\frac{dx}{\sqrt{\Omega_{\Lambda 0} + \Omega_{m0} x^3}} \,\; ,
\end{equation}
where $\Omega_{m0}$ and $\Omega_{\Lambda 0}$ are, respectively,
the matter and the cosmological density parameters, satisfying 
$\Omega_0 \equiv \Omega_{m0} + \Omega_{\Lambda 0} = 1$. Finally
we note that the horizon radius $d_{hor}$ is def\/ined as the 
limit of (\ref{redshift-dist}) when $z \to \infty$.

To close this section we remark that it is unlikely that 
astro-cosmological observations can f\/ix the density 
parameter $\Omega_0$ to be exactly equal to one. However, since 
the algebraic structure of the fundamental group of a constant 
curvature manifold is dif\/ferent for each
constant curvature $3$-geometry, the identif\/ication of the 
topology of space would unambiguously f\/ix the sign of the 
$3$-curvature. Moreover, if it turns out that $\Omega_0$ is 
exactly equal to one, topology seems to be the only way to 
precisely determine such a sharp value for 
$\Omega_0$~\cite{BerShvar}.

\section{Compact f\/lat space forms} 
\label{Flat3Man}

In this section we describe the classif\/ication of all f\/lat
compact $3$-manifolds (treated in full details in Appedix~A) 
and derive the explicit expressions for the corresponding 
injectivity radii $r_{inj}$. Further, we also present 
the formulae for the volumes of these $3$-manifolds, which are
obtained with some details in Appendix~B.    
The expressions for the injectivity radii are needed to build
the topological indicator $T_{inj}$ (Section~\ref{DetTop}),
which together with the expressions for the volumes are employed
in Section~\ref{Toys} to concretely discuss the detectability
of the topology of classes of compact f\/lat cosmological 
models.

The dif\/feomorphic classif\/ication of Euclidean $3$-dimensional 
space forms is well known~\cite{Wolf}. There are ten classes of 
compact Euclidean $3$-manifolds, six of which are orientable. 
Tables~\ref{Tb:OESF} and \ref{Tb:NOESF} give the dif\/feomorphism 
classes of orientable and non-orientable compact Euclidean 
$3$-dimensional space forms, respectively. In these tables, the 
triple $\{a,b,c\}$ is a set of three linearly independent 
vectors in Euclidean $3$-space (basis). An isometry in Euclidean 
$3$-space is denoted by $(A,a)$, where $a$ is a vector and $A$ is 
an orthogonal transformation, and the action of $(A,a)$ is given 
by    
\begin{equation}
\label{action}
(A,a) : x \mapsto Ax + a \; ,
\end{equation}
for any vector $x$. An isometry of the form $(I,a)$, where $I$ is 
the identity transformation, is written simply as $a$.

\begin{table}[t]
\begin{center}
\begin{tabular}{*{4}{|c}|} \hline & & & \\
Class & Generators of $\Gamma$ & $r_{inj}$ & Volume \\ 
& & & \\ \hline \hline
& & & \\ 
$\firstG$ & $a$, $b$, $c$ & $\frac{1}{2} |a|$ & $|a \times b 
\cdot c|$ \\ & & & \\ \hline
& & & \\ 
$\secondG$ & ($A_1$,$a$), $b$, $c$ & $\frac{1}{2} \min\{|a|,|b|\}$ 
& $|a| \, |b \times c|$ \\ & & & \\ \hline 
& & & \\ 
$\thirdG$ & ($B$,$a$), $b$, $c$ & $\frac{1}{2} \min\{|a|,|b|\}$ 
& $\frac{\sqrt{3}}{2} |a| \, |b|^2$ \\ & & & \\ \hline
& & & \\ 
$\fourthG$ & ($C$,$a$), $b$, $c$ & $\frac{1}{2} \min\{|a|,|b|\}$ 
& $|a| \, |b|^2$ \\ & & & \\ \hline
& & & \\ 
$\fifthG$ & ($D$,$a$), $b$, $c$ & $\frac{1}{2} \min\{|a|,|b|\}$ 
& $\frac{\sqrt{3}}{2} |a| \, |b|^2$ \\ & & & \\ \hline
& ($A_1$,$a$), & & \\
$\sixthG$ & ($A_2$,$b+c$), & $\frac{1}{2} \min\{|a|,|b|,|c|\}$ 
& $2 |a| \, |b| \, |c|$ \\ 
& ($A_3$,$a+b+c$) & & \\ \hline
\end{tabular}   %
\caption[Compact orientable $3$-dimensional Euclidean space 
forms.] {\label{Tb:OESF} \footnotesize Dif\/feomorphism classes of 
compact orientable $3$-dimensional Euclidean space forms. 
The f\/irst column contains Wolf's notation for each class. 
The second gives the generators of the corresponding 
covering groups.
The third column gives the injectivity radius, and the 
fourth  the volume. An isometry in Euclidean $3$-space 
is written as $(A,a)$, where $A$ is an orthogonal transformation 
and $a$ is a vector. The action of the isometry $(A,a)$ on 
Euclidean $3$-space is given by~(\ref{action}), while the orthogonal 
transformations in the second column are given by 
(\ref{Rot3}). From the isometric classif\/ication of f\/lat 
3-manifolds (Appendix~A) one has that for class $\firstG$ 
the vectors $a,b,c$ can always be ordered such that 
$|a| \leq |b| \leq |c|$. Further, for the class $\secondG$ 
the vectors $b$ and $c$ can be always be ordered such that 
$|b| \leq |c|$. Finally, for the classes $\thirdG$~--~$\fifthG$ 
one always has $|b| = |c|$. This makes apparent why the parameters 
$|b|$ and $|c|$ do not appear in the expression of $r_{inj}$ 
for the class $\firstG$, and also why $|c|$ does not appear 
in the expressions for $r_{inj}$ for the classes 
$\secondG - \fifthG$.}
\end{center}
\end{table}

The orthogonal transformations that appear in the classif\/ication 
of the Euclidean space forms take, in the basis $\{a,b,c\}$, the 
following matrix representations~\cite{Wolf}: 
\begin{eqnarray}
\label{Rot3}
A_1 = \left( \begin{array}{ccc}
               1 &  0 & 0 \\
               0 & -1 & 0 \\
               0 &  0 & -1
      \end{array} \right) \; , & 
A_2 = \left( \begin{array}{ccc}
              -1 & 0 & 0 \\
               0 & 1 & 0 \\
               0 & 0 & -1
      \end{array} \right) \; , & 
A_3 = \left( \begin{array}{ccc}
              -1 &  0 & 0 \\
               0 & -1 & 0 \\
               0 &  0 & 1
      \end{array} \right) \; , \nonumber \\ \\
B = \left( \begin{array}{ccc}
              1 & 0 & 0 \\
              0 & 0 & -1 \\
              0 & 1 & -1
    \end{array} \right) \; , & 
C = \left( \begin{array}{ccc}
              1 & 0 & 0 \\
              0 & 0 & -1 \\
              0 & 1 & 0
    \end{array} \right) \quad\mbox{and} & 
D = \left( \begin{array}{ccc}
              1 & 0 & 0 \\
              0 & 0 & -1 \\
              0 & 1 & 1
    \end{array} \right) \; , \nonumber
\end{eqnarray}
for the rotations, and 
\begin{equation}
\label{Ref3}
E = \left( \begin{array}{ccc}
               1 & 0 & 0 \\
               0 & 1 & 0 \\
               0 & 0 & -1
    \end{array} \right) \qquad \mbox{and} \qquad
F = \left( \begin{array}{ccc}
               1 & 0 &  2 \\
               0 & 1 &  1 \\
               0 & 0 & -1
    \end{array} \right) \, .
\end{equation}
for the ref\/lections.

\begin{table}[t]
\begin{center}
\begin{tabular}{*{4}{|c}|} \hline & & & \\
Class & Generators of $\Gamma$ & $r_{inj}$ & Volume \\ 
& & & \\ \hline \hline
& & & \\
$\firstB$ & ($E$,$a$), $b$, $c$ & $\frac{1}{2} \min\{|a|,|b|,|c|\}$ 
& $|a \times b| \, |c|$ \\ & & & \\ \hline
& & & \\
$\secondB$ & ($F$,$a$), $b$, $c$ & $\frac{1}{2} \min\{|a|,|b|,|c|\}$ 
& $|a \times b| \, |c - ( a + \frac{1}{2} b)|$ \\ & & & \\ \hline
& & & \\
$\thirdB$ & ($A_1$,$a$), ($E$,$b$), $c$ & $\frac{1}{2} 
\min\{|a|,|b|,|c|\}$ & $|a| \, |b| \, |c|$ \\ & & & \\ \hline
& & & \\
$\fourthB$ & ($A_1$,$a$), ($E$,$b+c$), $2c$ & 
$\frac{1}{2} \min\{|a|,|b|,2|c|\}$ & $2 \, |a| \, |b| \, |c|$ \\ 
& & & \\ \hline
\end{tabular}
\caption[Compact non-orientable $3$-dimensional Euclidean space 
forms.] {\label{Tb:NOESF} \footnotesize Dif\/feomorphism classes of 
compact non-orientable $3$-dimensional Euclidean space forms. The 
contents of the columns are of the same type as in Table 
\ref{Tb:OESF}. The orthogonal transformations in the second 
column are given by (\ref{Rot3}) and (\ref{Ref3}).}
\end{center}
\end{table}

Euclidean manifolds are not \emph{rigid}, in the sense that they 
can be deformed while still conserving its zero curvature at every 
point. As a consequence, manifolds in each class are topologically 
equivalent but can have dif\/ferent sizes, and even shapes, i.e. 
although dif\/feomorphic they may not be isometric. For example, a 
$3$-torus can be constructed by taking any parallelepiped and 
identifying opposite faces by translations.%
\footnote{Usually, it is considered a parallelepiped with mutually 
orthogonal faces (a brick), the simplest example being that of a 
cube. }
Note that parallelepipeds with dif\/ferent volumes give rise to 
non-isometric torii. Also stretching the parallelepiped in one or 
more directions while leaving the volume constant also gives rise 
to non-isometric torii. 

More exactly, the generators of the covering group of a torus form 
a basis in Euclidean $3$-space. Since the only condition for three 
vectors in Euclidean $3$-space to form a basis is to be linearly 
independent, no restriction is imposed on the lengths of these 
vectors, nor on their mutual orientations, i.e. the angles between 
them. These six parameters (three lengths and three angles) 
uniquely characterize a torus ``locally" in the $6$-dimensional 
parameter space formed by the lengths of the basis vectors and 
the angles between them. So, two torii with these six parameters 
approximately equal are non-isometric, but very similar in shape 
and size. However, this characterization is not unique. This can be 
seen by considering any basis $\{a,b,c\}$, and using it to construct 
another basis, say $\{a,a+b,c\}$, that generates the same fundamental 
group, and hence the same torus. Clearly the lengths of the vectors 
and the angles between them are very dif\/ferent for these two bases. 
Thus we have two dif\/ferent sets of these six parameters 
characterizing the same torus.

It is therefore clear that, to uniquely characterize isometrically 
a  torus, one has to perform some identif\/ications in the 
$6$-dimensional parameter space formed by the lengths of the basis 
vectors and the angles between them. This resulting quotient space 
is a kind of modular space for the torus, and uniquely gives the 
isometric classif\/ication for it. The isometric classif\/ication, 
i.e. the modular spaces, of Euclidean $3$-dimensional compact space 
forms is given with some details in Appendix~A, where we rectify 
some points of Wolf's classif\/ication~\cite{Wolf}.

A natural way to characterize the shape of compact manifolds is 
through the size of their closed geodesics. A suitable indicator 
is constructed using the injectivity radius def\/ined by 
(see~\cite{GRT1,GRT2})
\be
\label{rinj}
r_{inj} = \frac{1}{2} \min_{(g,x) \in \widetilde{\Gamma} \times P} 
\{\delta_g(x)\} \;,
\ee
where $\widetilde{\Gamma}$ denotes the covering group without 
the identity map, i.e. $\widetilde{\Gamma} = \Gamma \setminus 
\{id\}$, and $P$ is any fundamental polyhedron for $M$. The 
distance function $\delta_g(x)$ for a given isometry $g \in 
\Gamma$ is def\/ined by
\be
\label{dist-function}
\delta_g(x) = d(x,gx) \; ,
\ee
for all $x \in P$, where $d$ is the Euclidean metric. The distance 
function gives the length of the closed geodesic associated with 
the isometry $g \in \Gamma$ that passes through the projection of 
$x$ onto $M$. So, from~(\ref{rinj}) and~(\ref{dist-function}) one 
has that the injectivity radius is half of the length of the 
smallest closed geodesic in $M$, or equivalently, the radius of the 
smallest sphere inscribable in $M$.%
\footnote{Incidentally, as far as we are aware, the injectivity 
radius was introduced in Cosmic Topology by Sokolov and 
Shvartsman~\cite{SokShvar} who called it the ``minimum gluing 
parameter $l$''. It was def\/ined by saying  that ``at distances 
$r<l$ there isn't a single ghost''. 
In modern terms it means that any survey with depth smaller than 
the injectivity radius does not present multiple images nor 
pattern repetitions (see Section~\ref{DetTop}).}

In a globally homogeneous manifold, the distance function for 
any covering isometry $g$ is constant, and so is the length of the 
closed geodesic associated to $g$ and that passes through any $x 
\in M$. However, this is not the case in a locally, but not 
globally, homogeneous manifold, so the calculation of $r_{inj}$ 
for these cases requires some careful work.

To compute $r_{inj}$ it is convenient to choose a faithful 
representation of $M$. We take the Dirichlet domain $P$ of $\Gamma$ 
centered at the origin of Euclidean $3$-space. Consider the subset 
$\Delta \subset \widetilde{\Gamma}$ consisting on the isometries 
that transform $P$ to a neighbouring cell in the correspondent 
tessellation. Thus for all $g \in \Delta$, $gP \cap P$ is either a 
face, an edge, or a vertex of $P$. Now, for any $g = (A,a_g) \in 
\Delta$, the set of points that $A$ leaves unchanged (the axis of 
rotation if $\det(A)=1$, or the plane of ref\/lection if $\det(A) 
= -1$) passes through the origin and thus intersects $P$, so
\be  \label{mindelta}
\min_{x \in P} \{\delta_g(x)\} = |a_g^{\|}| \; ,
\ee
where $a_g^{\|}$ is the projection of $a_g$ onto the set of f\/ixed 
points of $A$. From eqs.~(\ref{rinj}) and~(\ref{mindelta})
and taking into account the above reasoning we have
\begin{eqnarray}
r_{inj} & = & \frac{1}{2} \min_{(g,x) \in \Delta \times P} 
\{\delta_g(x)\} \\ 
        & = & \frac{1}{2} \min_{g \in \Delta} \left\{\min_{x \in P} 
\{\delta_g(x)\}\right\} \\
        & = & \frac{1}{2} \min_{g \in \Delta} \{|a_g^{\|}|\} \; .
\end{eqnarray}

Analyzing separately each case from Tables~\ref{Tb:OESF} 
and~\ref{Tb:NOESF} one can compute the injectivity radius for 
any compact Euclidean $3$-dimensional space form. The results 
are exhibited in the third column of each table. 
Let us illustrate this procedure for the case of manifolds of 
class $\sixthG$. Firstly note that the matrices $A_1$, $A_2$ and 
$A_3$ satisfy the products $A_i A_j = A_k$, where the indices 
$i,j,k$ run in a cyclic order. Secondly note that the vectors 
$a$, $b$, and $c$, are parallel to the axes of rotation of 
$A_1$, $A_2$ and $A_3$ respectively. As a consequence, 
any isometry of the covering group of a manifold of class 
$\sixthG$ is of the form $\lambda = (A_i,u)$, where $i=1,2,3$ 
and $u$ is a linear combination of the vectors $a$, $b$, and $c$. 
Since $\lambda$ has no f\/ixed points, then the coef\/f\/icient 
of $a$ is a non-zero integer if $i=1$, and similarly for other 
values of $i$. It is now clear that $u^{\|}$ is a multiple of 
$a$, $b$, or $c$ for $i=1,2$ or 3 respectively. Thus the minimum 
length of a closed geodesic in a $\sixthG$ manifold is 
$\min\{|a|,|b|,|c|\}$.

To conclude this section note that the volumes of all closed 
Euclidean $3$-manifolds are listed in the fourth column of 
Tables~\ref{Tb:OESF} and~\ref{Tb:NOESF}. The relevant calculations 
are given in Appendix~B.

\section{Detectability problem in cosmic topology} 
\label{DetTop}

Regardless of our present-day inability to predict the 
topology of the universe, its detection and determination 
is ultimately expected to be an observational problem. 
At present it is becoming clear that the detection of 
a possible nontrivial topology of the universe 
may be a dif\/f\/icult problem to accomplish in view of the 
bounds on the cosmological parameters set by recent 
observations~\cite{GRT1}~--~\cite{Weeks2002}. Indeed, it was 
shown in~\cite{GRT1,GRT2} (see~\cite{EvLeLuUzWe} and~\cite{WeLeUz} 
for detectability of spherical spaces, and also~\cite{GRT3}and~%
\cite{Weeks2002} for the hyperbolic manifolds) that, if one uses 
pattern repetitions, increasing number of nearly f\/lat spherical and 
hyperbolic possible topologies for the universe become 
undetectable as $\Omega_0 \to 1$. It would appear at f\/irst 
sight that in the limiting case $\Omega_0 = 1$ the topology 
of such universes would def\/initely be undetectable. It 
turns out, however, that when $\Omega_0$ is exactly one the 
topological possibilities for the universe are completely 
dif\/ferent, and the detectability of cosmic topology may 
again become possible. In this section we shall brief\/ly 
review our approach to the detectability of cosmic topology 
in order to study explicitly in Section~\ref{Toys} the 
detectability of the topology of f\/lat universes.

For cosmological models with  compact spatial sections $M$ 
which have nontrivial topology it is clear that any attempt 
at the discovery of such a topology through observations must 
start with the comparison between the horizon radius and 
suitable characteristic sizes of the manifold $M$. We use the 
injectivity radius as a characteristic size of $M$. The ratio 
of the injectivity radius to the horizon radius,
\be 
\label{T_hor}
T_{inj}^{hor}=\frac{r_{inj}}{d_{hor}} \; ,
\ee
is very useful to identify cosmological models whose topology is 
undetectable through methods that rely on the existence of multiple 
images or pattern repetitions in any survey. In fact, for the case 
in which $T_{inj}^{hor} \geq 1$ the whole observable universe lies 
inside a fundamental polyhedron of $M$, no matter what is the 
location of the observer in the manifold (universe). In such cases 
no multiple images (or pattern repetitions) will arise from any 
survey. Thus, any method for the search of cosmic topology based on 
the existence of repeated patterns (multiple images) 
will fail --- the topology of the  universe is {\em definitely 
undetectable\/} in such cases. Note that this is nothing but 
a formal rewording of Sokolov and Shvartsman's statement quoted 
in footnote~2. 

We shall now discuss the detectability problem when we restrict the 
search to specif\/ic catalogues. There are basically three types of 
catalogues which can possibly be used in order to search for 
repeated patterns in the universe: namely, clusters of galaxies, 
containing clusters with redshifts of up to $z \approx 0.3$; 
active galactic nuclei (mainly QSO's and quasars), with a redshift 
cut-of\/f of $z_{max} \approx 4$; and maps of the CMBR with a 
redshift of $z \approx 10^3$. In this way, instead of $d_{hor}$ it 
is observationally more suitable to consider the largest distance 
$d_{obs} = d(z_{max})$ covered by a given survey, and def\/ine the 
indicator
\be 
\label{T_inj}
T_{inj}=\frac{r_{inj}}{d_{obs}} \; .
\ee
In this context, we shall refer to the region covered by a given 
survey by ``observed universe".

Now, if for a given catalogue $T_{inj} > 1$, every source in the 
survey is inside a fundamental polyhedron of $M$, no matter the 
location of the observer within the manifold. Actually, the whole 
``observed universe" lies inside a fundamental polyhedron of $M$. 
So, there are no multiple images in that survey any method for 
the search of cosmic topology based on the existence of repeated 
patterns will fail --- the topology of the universe is 
\emph{undetectable} with this specif\/ic survey. 
Thus, there emerges trivially from eq.~(\ref{T_inj}) and the 
expressions for the injectivity radii for all classes of f\/lat 
manifolds given in Section~\ref{Flat3Man} the undetectability 
conditions for each (topological) class of f\/lat universes. 
Equations~(\ref{UndRegion}) of Section~\ref{Toys} 
constitute an example of such conditions for the f\/lat classes 
$\thirdG$ and $\fifthG$ (see also~\cite{GRT4} for another explicit 
example). 

The bounds provided by recent cosmological 
observations~\cite{BooMax}~--~\cite{WMAP} can be used to identify f\/lat 
models having undetectable topologies, since an absolute lower bound 
of $r_{inj}$ for undetectability of f\/lat universes can be obtained 
by calculating the horizon radius corresponding to the limiting 
values of the density parameters. Indeed, for $\Omega_{m0}=0.3$ and 
$\Omega_{\Lambda 0}=0.7$, one obtains $d_{hor} \approx 10000 
h^{-1}\,Mpc$ from equation~(\ref{redshift-dist}). Thus, f\/lat 
universes for which $r_{inj} \geq 10000 h^{-1}\,Mpc$ have 
undetectable topologies.%
\footnote{Note that, e.g., for 
current catalogues of clusters ($z_{max}=0.3$), the undetectable 
f\/lat universes are now those for which $r_{inj} \geq 840 h^{-1} 
\,Mpc$, making clear that relative lower bounds arise when 
dif\/ferent catalogues are used.}

Some remarks are in order here. The indicator $T_{inj}$ is useful 
for the identif\/ication of cosmological models whose topology is 
undetectable by search methods based on the presumed existence of 
multiple images, for when $T_{inj} \geq 1$, the whole ``observed 
universe" lies inside a fundamental polyhedron of $M$. However, it 
should be noted that without further considerations, nothing can be 
said when $T_{inj} < 1$. In fact in this case, despite the depth of 
a given survey is larger than $r_{inj}$, it may be that, due to the 
location of the observer, the ``observed universe" would still be 
inside a fundamental polyhedron of $M$ making the topology 
undetectable. This is the case when the smallest closed geodesic 
that passes through the observer is larger than $2 d_{obs}$.

There is the case when $d_{obs}$ is larger than the length of the 
smallest closed geodesic that passes through the position of the 
observer, but not too much larger, so that only a small fraction of 
the "observed universe" contains multiple images. Current methods 
that look for multiple images are not sensitive enough to detect 
this small quantity of copies, so even in this case the topology of 
the universe would be in practice undetectable until suitable 
ref\/ined new methods are developed (and used). These and other 
points will be dealt with in future works.

The topology of a given cosmological model being undetectable for a 
given survey up to a depth $z_{max}$, clearly may be detectable by 
using a deeper survey. However, the deepest catalogue (survey) 
ever constructed will have $z_{max} < z_{SLS} \approx 10^3$. Thus 
the quotient~(\ref{T_inj}) computed with $z_{SLS}$ is a lower bound 
for the indicator $T_{inj}$. It turns out that, in practice, there 
is almost no dif\/ference if we push $z_{SLS}$ to inf\/inite, and 
take $T_{inj}^{hor}$ as the lower bound for $T_{inj}$ 
(see~\cite{GRT1,GRT2}).

\section{Detectability of f\/lat topologies: a case study} 
\label{Toys}

Since there is no natural unit of length in Euclidean geometry, 
there is no natural way to relate the typical lengths of the 
spatial sections of a f\/lat cosmological model with the density 
parameters within the General Relativity framework, as in the 
case of non-f\/lat cosmological models (see e.g.~\cite{BerShvar} 
and~\cite{GRT1}). So one should rely on other kinds of grounds to 
establish such a connection. Althouigh it is not our purpose 
here to discuss the origin and motivation of this sort of 
connection, it is interesting to concretely explore the 
consequences of one of such a hypothetical interrelation.
For this purpose we will construct some toy models based on a 
hypothesis f\/irst proposed by Bernshte\v{\i}n and 
Shvartsman~\cite{BerShvar}. A word of clarif\/ication is in order 
here. We emphasize that we are not claiming that this is a 
sound assumption for constructing realistic theoretical models 
for our universe. Instead, our intention here is to illustrate 
how a hypothesis like this, that may arrive from, e.g., a 
fundamental theory unifying elementary particles with gravity,%
\footnote{Note that we are not suggesting  nor conjecturing 
that the Bernshte\v{\i}n and Shvartsman assumption can arise from 
any unif\/ied theory of fundamental interactions.}
can be used to construct models which can be in principle 
confronted with cosmological observations.  

Thus, following Bernshte\v{\i}n and Shvartsman~\cite{BerShvar} we 
will suppose that the total number of baryons in our universe equals 
the reciprocal of the square of the gravitational f\/ine structure 
constant, 
\begin{equation}
\label{NumBar}
N = \alpha_{gr}^{-2} = \left(\frac{G m_p^2}{\hbar c}\, 
\right)^{\!\!-2} \approx \, 2.87 \cdot 10^{76} \;,
\end{equation}
where $G$ is the gravitational constant, here $h= 2\pi\,\hbar$ is 
Planck's constant and $m_p$ is the proton mass. This hypothesis 
enables one to construct cosmological models in which the volume of 
the universe is related to cosmological parameters and fundamental 
physical constants as follows. Since the total baryonic mass is 
$M_b=Nm_p$, the volume of our universe is 
\begin{equation}
\label{VolUniv1}
V = \frac{8 \pi}{3} \frac{GNm_p}{\Omega_b H_0^2} \; ,
\end{equation}
where we have used $\Omega_b = \frac{8 \pi G \rho_b}{3 H_0^2}$ for 
the baryon density parameter. Taking the current value for 
$\Omega_b h^2 = 0.03$  and $H_0 = 100 \, h \, km \, Mpc^{-1}/s$ 
(see, e.g.~\cite{Supernovae}), one obtains
\begin{equation}
\label{VolUniv2}
V \approx 2.9 \cdot 10^9 Mpc^3
\end{equation}
for the volume of the universe.

In order to illustrate the method of analysis of detectability 
of topology in these models, let us restrict to those ones whose 
spatial sections have a $\thirdG$ or $\fifthG$ topology. Note 
that the expressions for both $r_{inj}$ and the volumes of these 
two classes are identical, and moreover, the other classes can 
be dealt with in a similar way. Let us also consider that we are 
looking for multiple images with a catalogue of clusters of 
galaxies with redshift cut-of\/f $z_{max}=0.3$. Using the values 
$\Omega_{m0} = 0.3\,$ and $\Omega_{\Lambda 0} = 0.7$  
(see~\cite{BooMax,Supernovae}) for $h=0.7$ one obtains 
$d_{obs} = 1200 \, Mpc$. From Table \ref{Tb:OESF} we have that 
the region of undetectability is given by

\parbox{14.5cm}{
\begin{eqnarray*} 
|b| > 2\,d_{obs}  \quad & \mbox{if} &  \quad  |a| \geq |b| \;,\\
|a| > 2\,d_{obs} \quad & \mbox{if} &  \quad   |a|   <  |b| \;. 
\end{eqnarray*} }  \hfill 
\parbox{1cm}{
\begin{eqnarray} \label{UndRegion}  \end{eqnarray}  }

This region is shown in Figure~1, together with a curve of constant 
volume given by~(\ref{VolUniv2}). This f\/igure shows that the 
region of undetectability does not intersect the curve of constant 
volume, making clear that it may be possible to detect the topology 
in such universes. However, the possibility of detecting and even 
deciding the topology in these models clearly also depends on other 
factors such as the location of the observer in $M$, and the quality 
and reliability of the catalogues and methods used for the search of 
repeated patterns. We will deal with these problems in forthcoming 
articles.

\begin{figure}[!htb]
\centerline{\def\epsfsize#1#2{0.5#1}\epsffile{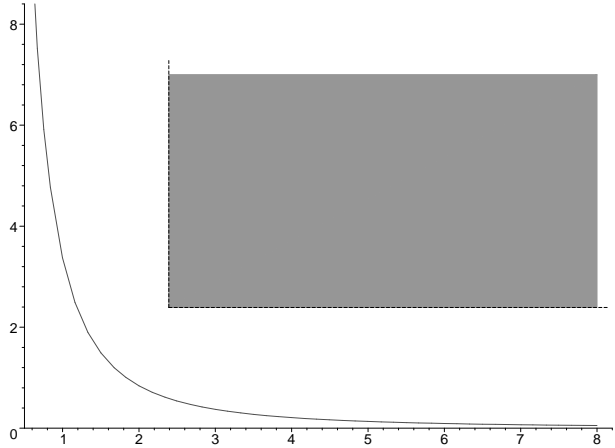}}
\caption{\label{fig=0.01}The undetectability region in the 
$(|a|\,,|b|)$ plane [def\/ined by~(\ref{UndRegion})] for 
universes with $\mathcal{G}_3$ and $\mathcal{G}_5$ topologies. 
Also a constant volume curve, whose expression is 
given in Table~1, for the value of the volume given 
by~(\ref{VolUniv2}). The vertical axis represents $|a|$, 
while the horizontal axis gives $|b|$.
Since the curve does not intersect the undetectability 
region it is potentially possible to detect the shape of 
such universes, with $\mathcal{G}_3$ and $\mathcal{G}_5$ 
topologies, using catalogues of clusters of galaxies 
($z_{max}=0.3$).} 
\end{figure}

It is interesting to mention that if one introduces a modif\/ied 
version of the Bernshte\v{\i}n-Shvartsman assumption in which $N$ 
equals the reciprocal of the square of a new gravitational f\/ine 
structure constant now given by 
\be
\alpha_{gr}^{-2} = \left(\frac{G m_p m_e}{\hbar c}\,\right)^{\!\!-2} 
\approx \, 9.68 \cdot 10^{82} \;,
\ee
where $m_e$ is the electron mass, following the above reasoning one 
obtains a f\/igure in which the curve of constant volume lies in 
the undetectability region, making clear that the topology of such 
f\/lat universes of classes $\thirdG$ and $\fifthG$ is undetectable 
even if CMBR is used. We shall not go into details of such a simple 
(and rather similar) calculations here for the sake of brevity.  

{}Furthermore, note that the above results are insensitive to 
substantial variations of the values of the cosmological density 
parameters $\Omega_{m0}$ and $\Omega_{\Lambda 0}$. In fact, even 
for the extreme cases of Einstein-de Sitter ($\Omega_{m0}=1$) and 
pure cosmological constant ($\Omega_{m0}=0$), the constant volume 
curve does not cross the undetectability region of Figure~1 if the 
Bernshte\v{\i}n-Shvartsman assumption is used, whereas if the above 
mentioned modif\/ied version of this assumption is employed one 
arrives at f\/lat universes with undetectable topologies. 

Alternatively, one can consider models in which the injectivity 
radius of the universe before inf\/lation was of the order of the 
Planck length. A simple calculation shows that to the topology be 
undetectable, inf\/lation should have expanded the universe more 
than a factor of $e^{135}\,$. Thus the detection of topology would 
set stringent constraints on inf\/lationary models.

\section{Conclusion and further remarks}  
\label{Concl}

It is becoming generally known that the detection of a possible 
nontrivial topology of the universe may be a dif\/f\/icult problem 
to accomplish in view of the bounds on the cosmological parameters 
set by recent observations which seem to indicate that we live in 
FLRW accelerating universe with nearly f\/lat spatial sections
Indeed, its has been shown in~\cite{GRT1}~--~\cite{WeLeUz} that if one 
uses pattern repetitions, an increasing number of nearly f\/lat 
spherical and hyperbolic possible topologies for the universe 
become undetectable as $\Omega_0 \to 1$. 

In this paper we have given a f\/irst prime of the dif\/feomorphic 
classif\/ication of compact f\/lat $3$-manifolds, and derive 
the expressions for both the injectivity radius $\,r_{inj}\,$ and 
the volume for each class of Euclidean compact $3$-manifolds. 

We have used the expression of the injectivity radius for the 
classes $\thirdG$ and $\fifthG$, and the topological indicator 
$T_{inj}$ in connection with 
Bernshte\v{\i}n-Shvartsman~\cite{BerShvar} assumption to illustrate
how one can materialize and quantify the problem of the detectability 
of f\/lat topologies. 
As a particular result we have shown that all models of these two 
classes of f\/lat universes that satisfy Bernshte\v{\i}n-Shvartsman 
hypothesis have a \emph{detectable} topology (at least in principle 
for some observers) if the existing catalogues of clusters of 
galaxies are used. We have also shown that a modif\/ied version of 
Bernshte\v{\i}n-Shvartsman assumption leads to f\/lat universes of 
classes $\thirdG$ and $\fifthG$ with \emph{undetectable} topologies.

In addition we have presented a brief discussion of the question of 
detectability of topology and ref\/ined upon some of the 
detectability aspects discussed in our previous 
articles~\cite{GRT1,GRT2}.


\vspace{3mm}
\section*{Acknowledgments}
We thank CLAF and CNPq for the grants under which this work 
was carried out.



\appendix
\vspace{5mm}

\section*{Appendices}

\section{Isometric classif\/ication of f\/lat 3-manifolds}
\label{Flat}

\begin{sloppypar}
In this appendix we shall present in full details the isometric 
classif\/ication of compact Euclidean $3$-dimensional space forms 
following and rectifying some points presented very brief\/ly by 
Wolf (see Lemma 3.5.11 and comments below on pages 123~--~124 
of~\cite{Wolf}). The isometric classif\/ication can be collected 
together in the following theorem:
\end{sloppypar}

\begin{theorem}
\label{Th:Isom-FE-E-3}
The isometric classif\/ication of Euclidean $3$-dimensional compact 
space forms is as follows:
\begin{enumerate}
\item The class $\mathcal{G}_1$ is parametrized by the equivalence 
classes $SL(3,Z) \cdot A(T^3) \cdot O(3)$, where $A(T^3)$ is the 
matrix whose rows are formed by the vectors of a base of 
Euclidean $3$-space that generates the covering group of the torus.
\item The class $\secondG$ is parametrized by the ordered triples 
$(|a|,r,w)$, where the pair $(r,w)$ parametrizes the bidimensional 
torus generated by the vectors $b$ and $c$.
\item The classes $\mathcal{G}_3$, $\mathcal{G}_4$ and 
$\mathcal{G}_5$ are parametrized by the ordered pairs 
$(|a|$,$|b|)$.
\item The classes $\sixthG$, $\mathcal{B}_3$ and $\mathcal{B}_4$ 
are parametrized by the ordered triples $(|a|$,$|b|$,$|c|)$.
\item The classes $\mathcal{B}_1$ and $\mathcal{B}_2$ are 
parametrized by the triples $(r,w,|c|)$, where the pair $(r,w)$ 
parametrizes the double covering of the modular space of the 
bidimensional torus generated by the vectors $a$ and $b$, which 
interchanges the generators of the lattice. Additionally, for the 
class $\secondB$, we have the constraint $|c| > |a + \frac{1}{2} 
b|$.
\end{enumerate}
\end{theorem}

\noindent
{\bf Proof.} We begin by constructing the modular space for the simplest 
classes, namely $\thirdG$, $\fourthG$ and $\fifthG$. Consider 
f\/irst the class $\thirdG$ and note that $Ba=a$, $Bb=c$, and 
$Bc=-\,(b+c)$. Since $B$ is an orthogonal transformation, it 
follows that $|b|=|c|=|b+c|$, and from 
\begin{eqnarray*}
\langle b,c \rangle & = & \langle Bb,Bc \rangle \\
                                    & = & - \, \langle c,b+c 
\rangle \\
                                    & = & - \langle b,c \rangle - 
|c|^2 \; ,
\end{eqnarray*}
one f\/inds that the angle between $b$ and $c$ is $\pi/3$. 
Furthermore, $a$ is orthogonal to $b$ and $c$. We have then that 
all the angles between the basis vectors are f\/ixed, and two 
lengths (those of $b$ and $c$) are equal, leaving just two free 
parameters, namely $|a|$ and $|b|$. It is now clear that the 
ordered pair $(|a|,|b|)$ parametrizes isometrically manifolds of 
class $\thirdG$.

Manifolds of classes $\fourthG$ and $\fifthG$ are analyzed in a 
similar way, by just looking at the action of the transformations 
$C$ and $D$ on the basis vectors, respectively. One obtains the 
following:
\begin{description}
\item[$\fourthG$.] The three basis vectors are mutually 
orthogonal, and $b$ and $c$ are of the same length. The ordered 
pair $(|a|,|b|)$ parametrizes isometrically this class.
\item[$\fifthG$.] The vectors $b$ and $c$ are of the same length 
and the angle between them is $\pi/6$, furthermore $a$ is 
orthogonal to both $b$ and $c$. Again, the ordered pair 
$(|a|,|b|)$ parametrizes isometrically this class.
\end{description}

Let us now turn our attention to manifolds with $3$-dimensional 
modular spaces, i.e. manifolds of classes $\sixthG$, $\thirdB$ 
and $\fourthB$. In all these cases the basis vectors are mutually 
orthogonal, as can be seen by analyzing the action of the 
transformations $A_1$, $A_2$, $A_3$ and $E$ on the vectors 
$a$, $b$ and $c$. Moreover, there is no restriction on the lengths 
of these vectors, so the parameter space is $3$-dimensional. In 
order to see that these classes are parametrized by the ordered 
triple $(|a|,|b|,|c|)$ note that the vector $a$ is the only one 
that enters as a translation along the axis of rotation of a screw 
motion. Also, for the classes $\thirdB$ and $\fourthB$, the vector 
$b$ is in the invariant subspace of the ref\/lection $E$, while $c$ 
is not. On the other hand, for the class $\sixthG$, the vector $b$ 
is in the invariant subspace of the rotation in the screw motion 
$(A_2,b+c)$, while $c$ does not have an analog property. Clearly, 
the vectors $a$, $b$ and $c$ play distinguished roles as 
translations in the generators of the corresponding covering 
groups.

We now brief\/ly review the isometric classif\/ication of the 
bidimensional torus in order to continue with the proof of theorem 
\ref{Th:Isom-FE-E-3}. The generators of the covering group of  
a 2-torus are two linearly independent vectors in the Euclidean 
plane, say $\xi = \{a,b\}$. These two vectors generate a lattice in 
the plane, namely
\[
\Lambda_{\xi} = \{n a + m b \; : \; n, m \;\mbox{are integers}\} 
\; .
\]
Actually, the lattice $\Lambda_{\xi}$ is the orbit of the origin 
under the action of the group generated by the basis $\xi$.

It is clear that two dif\/ferent bases give rise to equivalent 
lattices  if they are related by an orthogonal transformation of the 
plane, so one can always order the generators so that (i) $|a| \leq 
|b|$, (ii) $a$ lies along the positive direction of the $x-$axis, 
and (iii) $b$ is in the upper half plane with non-negative f\/irst 
component. Moreover, the bases $\{a,b\}$ and $\{a,n a + b\}$, with 
$n$ being a positive integer, give rise to the same lattice, so $b$ 
can be taken to be such that its projection to the $x-$axis lies 
between 0 and $|a|/2$. Thus a 2-torus is characterized isometrically 
by the pair $(r,w)$, where $r=|a|$, and $w$ is the complex number 
representing the same point in the plane as the vector $\frac{1}{r} 
\, b$. The parameter $w$ satisf\/ies the following conditions:
\begin{enumerate}
\item $|w| \geq 1$, and
\item $0 \leq \mbox{Re} \, w \leq 1/2$ .
\end{enumerate}

Next we study classes with 4-dimensional parameter spaces. Class 
$\secondG$ is the simplest, the vector $a$ is orthogonal to $b$ 
and $c$, and there is no restriction to the angle between these 
two vectors, nor to the lengths of the three basis vectors. Thus 
ordering the vectors $b$ and $c$ such that $|b| \leq |c|$ one 
concludes that the modular space of the class $\secondG$ is 
$(|a|,r,w)$, where $(r,w)$ is the modular space for the 
bidimensional torus generated by $b$ and $c$.

For the class $\firstB$ one notes that the vectors $a$ and $b$ 
are in the invariant subspace of the ref\/lection $E$, and $c$ 
is orthogonal to it. Since there is no restriction to the angle 
between $a$ and $b$, these two vectors form a bidimensional 
torus. However, one cannot simply set the condition $|a| \leq 
|b|$ without loss of generality, since the vectors $a$ and $b$ 
enter in a very dif\/ferent way in generating the covering 
group of the manifold, i.e. $b$ is a pure translation while $a$ 
enters as part of a glide ref\/lection. Hence we use a double 
covering of the modular space of the 2-torus generated by $a$ 
and $b$ to parametrize the class $\firstB$, one sheet 
parametrizing the case $|a| \leq |b|$, the other parametrizing 
the case $|b| \leq |a|$. The class $\secondB$ is similar to the 
class $\firstB$, except that $c$ is not orthogonal to the plane 
formed by $a$ and $b$, but has a projection $(a+ \frac{1}{2} 
b)$ on this plane. This is the origin of the constraint $|c| > 
|a+ \frac{1}{2} b|$.

Finally let us consider the class $\firstG$, or the $3$-dimensional 
torus. The isometric classif\/ication of the torus has been 
partially discussed in Section~\ref{Flat3Man}. There, we have seen 
that bases related by orthogonal transformations give rise to 
isometric torii, and furthermore, one has to perform additional 
identif\/ications on the set of bases of $3$-dimensional Euclidean 
space to give a unique characterization of isometric torii. To put 
this in mathematical (and useful) language, let us f\/irst choose 
as a reference basis the canonical one ($\hat{\imath} = (1,0,0)$, 
$\hat{\jmath} = (0,1,0)$, and $\hat{k} = (0,0,1)$) to write the 
components of vectors and transformations. Now, let us write a 
basis in Euclidean $3$-space as a square matrix whose rows are the 
components of the vectors of the basis. If the basis $\xi = 
\{e_1,e_2,e_3\}$ is related to the basis $\eta = \{f_1,f_2,f_3\}$ 
by an orthogonal transformation, one can write
\[
\eta = \xi B \; ,
\]
where $B \in O(3)$. In this case, the bases $\xi$ and $\eta$ 
generate two isometric torii.

A lattice in Euclidean $3$-space is the orbit of the origin by the 
action of a group generated by three linearly independent vectors. 
Thus if $\xi = \{e_1,e_2,e_3\}$, the lattice generated by $\xi$ 
is the set 
\[
\Lambda_{\xi} = \{n_1 e_1 + n_2 e_2 + n_3 e_3 \; : \; n_1,n_2,n_3 \; 
\mbox{are integers}\} \; .
\]
Now, two bases $\xi$ and $\eta$ may produce the same lattice, and 
hence generate the same torus. In fact, consider  the group 
$SL(3,Z)$ formed by the square matrices of order three with integer 
entries and determinant unity. Any two bases related by a matrix 
$A \in SL(3,Z)$ generates the same lattice, for if $\eta = A \xi$, 
each $f_i$ is an integer linear combination of $\xi$, and thus 
$\Lambda_{\eta} \subseteq  \Lambda_{\xi}$. On the other side, 
$\xi = A^{-1} \eta$, thus $\Lambda_{\xi} \subseteq \Lambda_{\eta}$, 
and so $\Lambda_{\xi} = \Lambda_{\eta}$. We have that two bases 
$\xi$ and $\eta$ generate the same torus if there exist $A \in 
SL(3,Z)$ and $B \in O(3)$ such that $\eta = A \xi B$.

\section{Volumes of f\/lat 3-manifolds}
\label{Volumes}

In this appendix we show how to calculate the volumes of compact 
f\/lat $3$-manifolds in terms of their modular spaces. The 
calculations are based on the following simple observations:
\begin{enumerate}
\item The volume of a compact manifold of constant curvature 
equals the volume of any of its fundamental domains.
\item If $\Gamma$ is the covering group of a compact f\/lat 
$3$-dimensional manifold in the form given in Tables \ref{Tb:OESF} 
and~\ref{Tb:NOESF}, the orbit of the origin by $\Gamma$ is a 
lattice, and a fundamental domain of this lattice is a 
fundamental domain of $\Gamma$.
\item A fundamental domain of the lattice generated by the base 
$\xi = \{a,b,c\}$ is the parallelepiped naturally constructed 
with these vectors, and thus its volume is $|a \cdot b \times c|$.
\end{enumerate}

Now, in terms of the correspondent modular spaces given in 
Theorem~\ref{Th:Isom-FE-E-3}, the orbit of the origin by the 
covering group of a manifold of any of the classes 
$\firstG$--$\,\fifthG$ or $\firstB$--$\,\thirdB$, is the lattice 
generated by the vectors $a$, $b$ and $c$. While the orbit of 
the origin by the covering group of a manifold of class $\sixthG$ 
is a lattice generated by the vectors $a$, $b+c$ and $b-c$. 
Finally, note that a manifold of class $\fourthB$ is a double 
covering of the manifold of class $\thirdB$ with the same 
parameters, hence its volume doubles that of the corresponding 
manifold of class $\thirdB$.

The results of this appendix are listed in the fourth column 
of Tables~\ref{Tb:OESF} and~\ref{Tb:NOESF}.


%
%
\end{document}